\documentclass{emulateapj}
\usepackage{graphics}
\usepackage{epsfig}
\usepackage{natbib}
\shorttitle{Diffuse \ion{Si}{2}$^{*}$ Emission Around KISSR242}
\shortauthors{France et al.}

\begin{document}


\title{Diffuse Far-UV Line Emission from the Low-Redshift Lyman Break Galaxy Analog KISSR242.\altaffilmark{1}}


%\author{Kevin France\altaffilmark{2}, Nicholas Nell, James C. Green, Brian A. Keeney}
\author{Kevin France\altaffilmark{2}, Nicholas Nell, James C. Green}

\affil{Center for Astrophysics and Space Astronomy, 389 UCB, University of Colorado, 
Boulder, CO 80309}

%\and

\author{Claus Leitherer}
\affil{Space Telescope Science Institute, 3700 San Martin Drive, Baltimore, MD 21218}
    

%%%These are for extra affiliations at the bottom of page
\altaffiltext{1}{Based on observations made with the NASA/ESA $Hubble$~S$pace$~$Telescope$, obtained from the data archive at the Space Telescope Science Institute. STScI is operated by the Association of Universities for Research in Astronomy, Inc. under NASA contract NAS 5-26555.}

\altaffiltext{2}{kevin.france@colorado.edu}
%\altaffiltext{3}{Exoplanets and Stellar Astrophysics Laboratory, 
%	NASA Goddard Space Flight Center, Greenbelt, MD 20771; akir@milkyway.gsfc.nasa.gov}
%\altaffiltext{4}{Department of Physics and Astronomy, Johns Hopkins University,
%   Baltimore, MD 21218; roxana@pha.jhu.edu, pdf@pha.jhu.edu}
%\altaffiltext{5}{McDonald Observatory, University of Texas at Austin, Austin, TX 78712; 
%sredfield@astro.as.utexas.edu}


\begin{abstract}

We present new ultraviolet (UV) observations of the luminous compact blue galaxy KISSR242, obtained with the {\it Hubble Space Telescope}-Cosmic Origins Spectrograph ($HST$-COS).  We identify multiple resolved sub-arcsecond near-UV sources within the 
COS aperture. The far-UV spectroscopic data show strong outflow absorption lines, consistent with feedback processes related to an episode of massive star-formation.  \ion{O}{1}, \ion{C}{2}, and \ion{Si}{2}~--~\ion{Si}{4} are observed with a mean outflow velocity 
$\langle v_{out} \rangle$ = -60 km s$^{-1}$.  We also detect faint fine-structure emission lines of singly ionized silicon for the first time in a low-redshift starburst galaxy.  These emissions have been seen previously in deep Lyman break galaxy surveys at $z$~$\sim$~3. 
The \ion{Si}{2}$^{*}$ lines are at the galaxy rest velocity, and they exhibit a quantitatively different line profile from the absorption features.  These lines have a width of $\approx$~75 km s$^{-1}$, too broad for point-like emission sources such 
as the \ion{H}{2} regions surrounding individual star clusters.    
The size of the \ion{Si}{2}$^{*}$ emitting region is estimated to be $\approx$~250~pc.
We discuss the possibility of this emission arising in overlapping super star cluster \ion{H}{2} regions, but find this explanation to be unlikely in light of existing far-UV observations of local star-forming galaxies.  We suggest that the observed \ion{Si}{2}$^{*}$ emission originates in a diffuse warm halo populated by interstellar gas driven out by intense star-formation and/or accreted during a recent interaction that may be fueling the present starburst episode in KISSR242.

\end{abstract}



\clearpage

%%%%%%%%%%%%%%%%%%%%%%%%%%%START OF THE PAPER%%%%%%%%%%%%%%%%%%%%%%%

\section{Introduction}

Models of sustained star-formation predict an evolving relationship
between gas loss and gas accretion in galactic disks and halos~\citep{somerville99}.  Gas is lost through `feedback' processes such as supernova explosions and winds from massive star-forming regions.  Depending on the mechanical energy input into this outflowing material, it may escape into the intergalactic medium (IGM; Ferrara \& Tolstoy 2000), thereby enriching the IGM with the metal systems observed along quasar sightlines~\citep{danforth08,tripp08}.~\nocite{ferrara00}  Alternatively, this material may be re-accreted (`recycled'; Oppenheimer et al. 2010) onto the galaxy, providing fuel for future generations of star-formation.~\nocite{oppenheimer10}  Additional possible sources of the baryonic raw material needed for continued star-formation are the extended dark matter halos surrounding galaxies (and clusters), the intergalactic medium, and other galaxies encountered during an interaction.  
% In the first scenario, the virialized hot halo cools (hot mode accretion) allowing this mostly primordial gas to be accreted~\citep{rees77}.  
% The second scenario suggests that baryons residing in intergalactic filaments can be efficiently funneled into galaxies 
% (cold mode accretion; Keres et al. 2005; Dekel \& Birnboim 2006) triggering episodes of star-formation.~\nocite{keres05,dekel06}  
Intense star-formation observed in mergers (Zhang et al. 2010 and references therein) provides evidence that interactions are a means of acquiring both primordial and enriched gas.~\nocite{zhang10}  

Luminous compact blue galaxies (LCBGs) are relatively low-mass galaxies with high surface brightness owing to their high rates of star-formation~\citep{guzman03}.  This star-formation may be related to a previous gas accretion event, and is shown to be driving large scale outflows (i.e. feedback).  These objects have a relatively low dust content, and as a result, the rest-frame ultraviolet (UV) light from the young OB stars in these galaxies dominate their spectral energy distributions.  LCBGs are candidates for local analogs to the well-studied Lyman break galaxy (LBG) population, although they may be part of the lower mass end of the LBG analog distribution~\citep{guzman03,overzier09}.
The LBG classification is derived from the identification method which relies on a non-
detection of Lyman continuum emission ($\lambda_{rest}$~$<~$~912 \AA) in multi-band 
imaging surveys.  
Low-$z$ UV observations require a space-based platform, which severely limits the ability of observers to carry out surveys of these objects.  As a consequence, much more is known about these galaxies at redshifts~$\gtrsim$~3 when the rest-frame UV shifts into the optical and these objects can be studied by wide-field imaging and spectroscopic surveys employing 10-m class telescopes~\citep{steidel01,shapley03,stark10}.   

\citet{shapley03} presented a comprehensive survey of approximately 1000 LBGs, 
from which high quality composite spectra were created.  In addition to the wealth of stellar diagnostics accessible in the rest-frame UV (1000~$\lesssim$~$\lambda$~$\lesssim$~2000~\AA), moderate (\ion{C}{4}, \ion{Si}{4}; $T$~$\sim$~10$^{5}$ K) and low (\ion{O}{1}, \ion{C}{2}, \ion{Si}{2}; 1~--~3~$\times$~10$^{4}$ K) ionization outflow was detected with velocities of -(100~--~200) km s$^{-1}$.  The composite spectra also showed weak emission from \ion{Si}{2}$^{*}$ fine-structure lines.  These lines were not observed in individual galaxies, however they have been observed in the highly lensed $z$~=~2.73 LBG MS1512-cB58~\citep{pettini00,pettini02}, in several of the LBGs observed as part of the Lyman continuum search described by~\citet{shapley06}, and recently in deep observations of the $z$~=~2.3 $L^{*}$ galaxy Q2343-BX418~\citep{erb10}.  Interestingly, neither these silicon lines nor other similar low-ionization tracers have been detected in UV spectroscopic samples of star-forming galaxies in the low-$z$ universe~\citep{kinney93,grimes09,leitherer10}.  The physical state and geometric location of these fine-structure line emitting regions remains unexplained.  We have detected a similar \ion{Si}{2}$^{*}$ region in the LCBG KISSR242 ($z$~=~0.037813), using new observations from the $Hubble$~$Space$~$Telescope$-Cosmic Origins Spectrograph ($HST$-COS).  In this paper, we present initial evidence that this \ion{Si}{2}$^{*}$ emitting region is part of a diffuse low-ionization halo.  We favor a scenario where this halo is populated by outflowing interstellar gas or accreted as part of the stream that may be fueling the current episode of star-formation in KISSR242.  

\begin{figure}
\begin{center}
\hspace{+0.0in}
\epsfig{figure=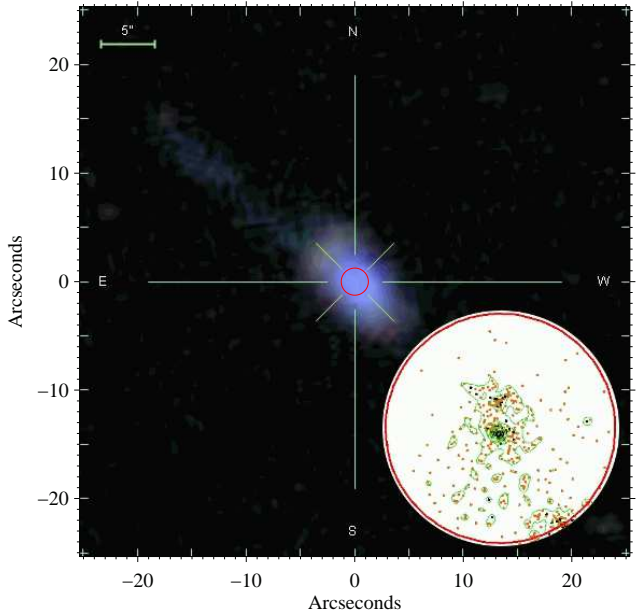,width=3.6in,angle=90}
\caption{\label{cosovly} SDSS image of KISSR242.  A faint blue stream is observed to the northeast of the main galaxy.   
The COS aperture is shown overlaid at the origin (R.A. = 13$^{\mathrm h}$ 16$^{\mathrm m}$ 03.90$^{\mathrm s}$, Dec. = +29\arcdeg\ 22\arcmin\ 53.8\arcsec; J2000) in red.  The inset to the southwest shows the COS near-UV target acquisition image.  Note that there are several sources in this short (2.0s) image, however the near-UV emission (individual counts are shown as the orange dots) are not filling the COS aperture.
Contours are overplotted in green to illustrate the central concentration.
 }
\end{center}
\end{figure}

%%%%%%%%%%%%%%%%%%%%%%%%%%%%%%%%%%%%%%%%%%%%%%%%%%%%%%%%%%%%%%%%%%%%%%%%%%%%%%%%
\section{$HST$-COS Observations}

We used the short-wavelength, medium resolution far-UV mode (G130M) of COS 
to observe KISSR242 on 26 and 27 December 2009.  The observations have a total exposure time of $T_{exp}$~$\approx$~2222s.  
Two central wavelengths were used ($\lambda$1291 and $\lambda$1318) at the default focal plane offset position (FP-POS~=~3) in order to provide continuous spectral coverage while minimizing the impact of microchannel plate detector fixed pattern noise and spacecraft overhead.  
%The dataset identifiers for the two exposures are {\tt lb6202010} and {\tt lb6202020}. 
All observations were centered on KISSR242 (R.A. = 13$^{\mathrm h}$ 16$^{\mathrm m}$ 03.90$^{\mathrm s}$, Dec. = +29\arcdeg\ 22\arcmin\ 53.8\arcsec; J2000) and COS performed an NUV imaging target acquisition through the primary science aperture.  This combination of grating settings covers the 1138~$\lesssim$~$\lambda$~$\lesssim$~1457~\AA\ bandpass. 
Figure 1 shows a Sloan Digital Sky Survey (SDSS) image of KISSR242, with the location of the 2.5\arcsec\ diameter COS aperture overlaid.
The inset in Figure 1 the NUV target acquisition image obtained immediately preceding the spectroscopic observations.
The data were reprocessed with  CALCOS v2.11k2, and combined with the custom IDL coaddition procedure described in Danforth et al. (2010).~\nocite{danforth10}

\begin{figure*}[t]
\begin{center}
\hspace{+0.0in}
\epsfig{figure=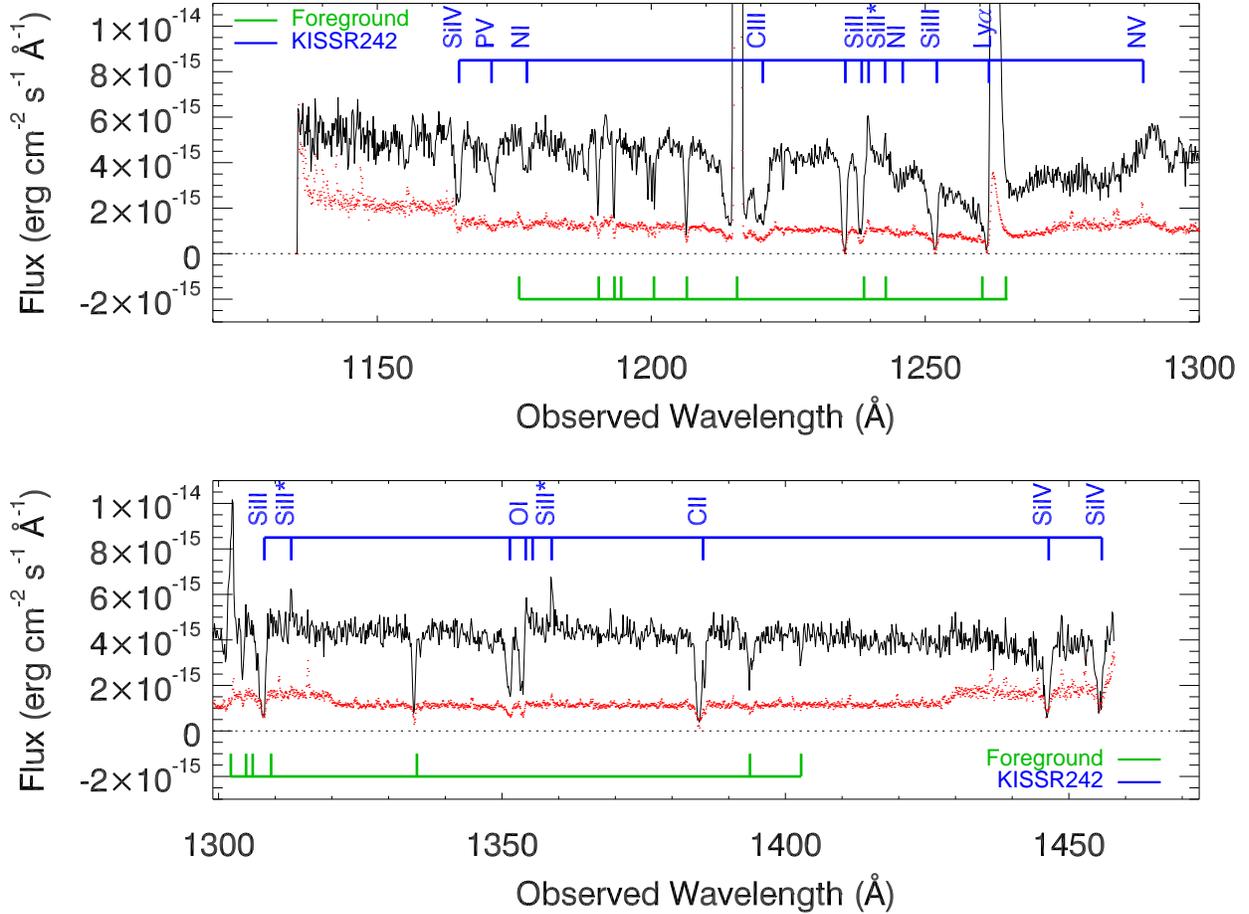,width=5.0in,angle=90}
\caption{\label{cosspec} $HST$-COS spectra of KISSR242, smoothed by two resolution elements for display purposes.  The data are shown in black, with the error vector overplotted as the dotted red line.  Abrupt changes in the error level are due to the non-uniform exposure time coverage associated with observations at multiple central wavelengths ($\lambda$1291 and $\lambda$1318).  Foreground absorbers are identified with green tickmarks and the emission and absorption features at the redshift of KISSR242 are labeled in blue.  
 }
\end{center}
\end{figure*}

%%%%%%%%%%%%%%%%%%%%%%%%%%%%%%%%%%%%%%%%%%%%%%%%%%%%%%%%%%
\section{Analysis} 

\subsection{Line Fitting and Galactic ISM}

Figure 2 displays the full, coadded COS G130M spectrum of KISSR242.  Numerous emission and absorption features are present in the spectrum, both from the Milky Way and in the environment of KISSR242.  Figure 2 identifies the Galactic and 
extragalactic features with green (Milky Way) and blue (KISSR242) tickmarks. 
All of the features were measured using custom IDL spectral line-fitting software developed by the COS Science Team.  This multiple-line fitting routine uses the appropriate wavelength-dependent COS line-spread function (LSF\footnote{The COS LSF experiences a wavelength dependent non-Gaussianity due to the introduction of mid-frequency wave-front errors produced by the polishing errors on the $HST$ primary and secondary mirrors; {\tt http://www.stsci.edu/hst/cos/documents/isrs/}}) to return the underlying Gaussian line-shape parameters.  We used the observed wavelengths to identify Galactic absorption from \ion{Si}{2}, \ion{N}{1}, \ion{Si}{3}, \ion{H}{1}, \ion{O}{1}, \ion{C}{2}, and \ion{Si}{4}.  
These lines are not of interest in the present work, and we identify them simply to remove any confusion introduced when analyzing the spectrum of KISSR242.  

\begin{deluxetable}{lcccc}[b]
\tabletypesize{\footnotesize}
\tablecaption{KISSR242 Absorption Lines\tablenotemark{a}}
\tablewidth{0pt}
\tablehead{
\colhead{Species} & \colhead{$\lambda_{rest}$\tablenotemark{b}} & \colhead{$\lambda_{lab}$}
& \colhead{FWHM}  &  \colhead{$v_{hel}$} \\ 
     & (\AA) & (\AA) &  (km s$^{-1}$) & (km s$^{-1}$) }
%\tableline
\startdata
\ion{Si}{4}       & 1122.30 & 1122.49   & 275 $\pm$ 14 & -51 $\pm$ 11 \\ 
\ion{Si}{2}       & 1190.22 & 1190.42  & 267 $\pm$ 6  & -48 $\pm$ 10 \\ 
\ion{Si}{2}       & 1193.07 & 1193.29  & 265 $\pm$ 7  & -55 $\pm$ 10 \\ 
\ion{Si}{3}       & 1206.24 & 1206.50     & 218 $\pm$ 15 & -64 $\pm$ 12 \\ 
\ion{Si}{2}       & 1260.05 & 1260.42  & 319 $\pm$ 9  & -88 $\pm$ 11 \\ 
\ion{O}{1}        & 1302.05 & 1302.17    & 215 $\pm$ 9  & -29 $\pm$ 10 \\ 
\ion{C}{2}        & 1334.28 & 1334.53  & 262 $\pm$ 6  & -56 $\pm$ 10 \\ 
\ion{Si}{4}       & 1393.43 & 1393.76   & 250 $\pm$ 12 & -69 $\pm$ 11 \\ 
\ion{Si}{4}       & 1402.39 & 1402.77    & 270 $\pm$ 15 & -82 $\pm$ 11 \\
 \enddata
%% NOTES IN TABLE
\tablenotetext{a}{Blended absorption lines are not displayed.} 
\tablenotetext{b}{Calculated assuming the SDSS spectroscopic redshift $z$~=~0.037813~$\pm$~0.000028.}
\end{deluxetable}

\subsection{KISSR242 Absorption and Emission Lines }

The spectra were shifted into the rest frame of KISSR242 ($z$~=~0.037813~$\pm$~0.000028; SDSS DR6 spectroscopic redshift), and absorption and emission lines associated with the galaxy were fit.  A line-identification analysis finds four classes of spectral features in the far-UV spectra of KISSR242: 1) photospheric absorption from young, massive stars; 2) interstellar absorption that is related to the star-formation driven outflow; 3) strong Ly$\alpha$ emission and absorption; 4) emission from fine-structure lines of singly ionized silicon.
We observe absorption in the intermediate to high excitation lines of \ion{P}{5} $\lambda$1118, 1128, \ion{Si}{4} $\lambda$1122, 1128, the \ion{C}{3} $\lambda$1175 multiplet, and \ion{Si}{4} $\lambda$1394, 1403 that are characteristic of the atmospheres and winds of massive stars~\citep{walborn02}.  While all of these features are temperature sensitive, the presence of \ion{P}{5} in the observations confirms the existence of young O-stars in KISSR242~\citep{pellerin02}, as expected for a LCBG.  Unfortunately, at the effective resolution of these observations (\S3.3), line-blending and low-S/N prevented us from measuring reliable line parameters for several of these species.  

We observe strong blue-shifted absorption from KISSR242. We observe interstellar absorption from \ion{N}{1} $\lambda$1135, 1200, \ion{Si}{2} $\lambda$1190, 1193, 1260, 1304, \ion{Si}{3} $\lambda$1206, \ion{O}{1} $\lambda$1302, 1304, \ion{C}{2} $\lambda$1334, 1335, and \ion{Si}{4} $\lambda$1394, 1403 most likely has both stellar and interstellar components.  Blends also complicate several of these lines, and we present lines that could be cleanly measured in Table 1.  The mean outflow velocity is $\langle v_{out} \rangle$~=~-60~$\pm$~18 km s$^{-1}$, with an average line width of 260~$\pm$~31 km s$^{-1}$ (however, see \S3.3).  The outflow velocity for KISSR242 is essentially identical to the average \ion{C}{2} outflow velocity observed in local starbursts, $\langle v_{sb} \rangle$~$\approx$~-65 km s$^{-1}$~\citep{grimes09}.  This outflow velocity is somewhat smaller than the average multi-species outflow velocity for LBGs at $z$~$\sim$~3, $\langle v_{LBG} \rangle$~$\approx$~-127 km s$^{-1}$~\citep{shapley03}, although it has been shown that the mean outflow velocity may not be a reliable measure of the true outflow distribution~\citep{steidel10}.  The Ly$\alpha$ absorption and emission lines suggest a multi-component structure.  A comprehensive analysis of these lines will be presented in the context of the larger COS starburst survey.

Finally, we observe fine-structure emission from singly-ionized silicon at the redshift of KISSR242.  Our confidence in this line-identification is strengthened by the firm detection of four lines from these fine structure states ($\lambda_{lab}$ = 1194.50, 1197.39, 1264.73, and 1309.27 \AA) as well as the appearance of these lines in the composite and individual spectra of $z$~$\sim$~3 LBGs~\citep{pettini00,shapley03,shapley06}.  
Unlike the stellar and interstellar lines, the \ion{Si}{2}$^{*}$ emission lines are found to be at rest with respect to the systemic velocity (as measured from the optical wavelength nebular emission lines) of the galaxy.  We find that the average velocity of the fine-structure emission is $\langle v_{emis} \rangle$~=~-4~$\pm$~10 km s$^{-1}$.  Observations of these lines at $z$~$\sim$~3 find \ion{Si}{2}$^{*}$ 
velocities of order +100 km s$^{-1}$, significantly different than the zero velocity emission we detect in KISSR242. The lines have an average width of 75~$\pm$~13 km s$^{-1}$, $\sim$~3.5 times narrower than the observed outflowing absorption (\S3.3).  
Figure 3 shows a select spectral window of the COS data of KISSR242, highlighting a region of both outflow absorption and the emission from \ion{Si}{2}.  The observed emission features (other than Ly$\alpha$\footnote{We note that there is also a broad (FWHM~$\approx$~630 km s$^{-1}$) emission feature observed at 1291.30~\AA.  One possibility for this feature is redshifted \ion{N}{5} $\lambda\lambda$1238, 1242, although we cannot make a conclusive identification.}) are quantified in Table 2.  

\subsection{KISSR242 Line Widths~--~The Effect of Aperture Filling Fraction}

Extended sources complicate the interpretation of line widths observed by slitless spectrographs.  In light of the comparison of the COS aperture size with the optical image of KISSR242 presented in Figure 1, we suggest that the absorption lines may be intrinsically narrower than $\langle FWHM_{out} \rangle$~=~260 km s$^{-1}$.  In this scenario, an aperture filling fraction of order unity imposes an artificial minimum to the observed absorption line widths.  As a check on this hypothesis, we have analyzed the cross-dispersion profiles in the raw two-dimensional COS spectrograms, confirming that the far-UV continuum source size (evaluated at $\lambda_{obs}$~$\approx$~1280~$\pm$~5~\AA) is 2.0~$\pm$~0.3\arcsec, mostly filling the 2.5\arcsec diameter COS aperture.  The observed absorption line widths in KISSR242 agree with this interpretation, and are consistent with filled aperture resolving powers in the range 1500~$\lesssim$~$R$~$\lesssim$~1000 (200~$\lesssim$~FWHM$_{out}$~$\lesssim$~300 km s$^{-1}$; Table 1).  These lines show no trend towards enhanced line width with ionization state, and are consistent with the filled-aperture COS observations presented by~\citet{france09}.  

Interestingly, the near-UV target acquisition image shown inset in Figure 1 suggests that the morphology is one of resolved individual sources in the $\sim$~2000~--~3000~\AA\ imaging band.  There are three main sources in the near-UV image: 1) the main central source with angular diameter $\approx$~0.11\arcsec, 2) the northern central source, at an angular separation from the main central source of 0.33\arcsec, and 3) a southwestern source at the edge of the COS aperture ($\Delta\theta$~$\approx$~1.23\arcsec).  
The differences between far-UV and near-UV source sizes is surprising, and we suggest two possible mechanisms that might explain the observation.  The first explanation invokes extended star-formation in the outer regions of KISSR242 that contributes strongly to the far-UV emission, while the older stars that dominate the near-UV emission are concentrated into a nuclear structure.  This could be analogous to the ``XUV disks'' found in more massive local starburst galaxies~\citep{thilker05,thilker07}.  Alternatively, if a significant amount of interstellar dust is present, the strong forward scattering of grains at far-UV wavelengths~\citep{burgh02,draine03} could act to diffuse the far-UV starlight observed towards KISSR242 while having relatively little impact on the near-UV light from the same stars.  Further investigation with $HST$ imaging at near- and far-UV wavelengths would be helpful.

The \ion{Si}{2}$^{*}$ emission we observe is significantly broader than both the 17 km s$^{-1}$ COS spectral resolution element and the thermal width expected for collisionally ionized silicon ($T_{SiII}$~$\sim$~3~$\times$~10$^{4}$ K; $\Delta$$v_{therm}$~$\sim$~5 km s$^{-1}$) in this state~\citep{dere09}.  The breadth of the observed emission lines argues against a physical picture where the fine-structure lines are simply nebular in nature, emitted from an individual \ion{H}{2} region around the central source.  Conversely, the lines are too narrow to be a source that is uniformly filling the COS aperture.   The 75 km s$^{-1}$ width of the fine-structure lines in KISSR242 correspond to an effective resolving power of $R$~$\sim$~4000.  Using a ray-trace model of the COS far-UV optical path, we estimate that this corresponds to a source angular diameter of $\theta_{SiII}$~$\sim$~0.34~$\pm$~0.06\arcsec. This calculation assumes a uniform emission source, the situation becomes less clear for a clumpy emitting geometry.   In the following section, we use this information about the spatial distribution of \ion{Si}{2}$^{*}$ to suggest possible sources for this enigmatic emission.

\begin{figure}
\begin{center}
\hspace{+0.0in}
\epsfig{figure=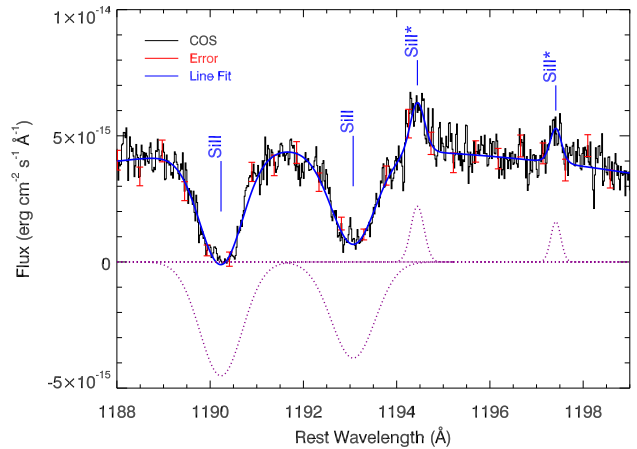,width=2.6in,angle=90}
\caption{\label{cosfit} Zoomed view of the \ion{Si}{2} absorption ($\lambda$1190, 1193; Table 1) and \ion{Si}{2}$^{*}$ emission ($\lambda$1194, 1197; Table 2) from KISSR242.  The multi-component fit is shown in blue, with the individual Gaussian absorption and emission components overplotted as the dark magenta dotted lines. 
 }
\end{center}
\end{figure}

%%%%%%%%%%%%%%%%%%%%%%%%%%%%%%%%%%%%%%%%%%%%%%%%%%%%%%%%%%%%%%
%\section{Discussion}

\section{Fine-Structure Emission in KISSR242: \ion{H}{2} Regions, Outflow or Accreted Halo? }

We present two possible scenarios for the distribution of line-emitting gas in the KISSR242 system: 1) Overlapping \ion{H}{2} regions from nuclear super star clusters (SSCs) and 2) a static distribution of halo gas populated by starburst-driven outflow or the recent interaction with another galaxy.  First, we will define a rough size-scale for the observed emission.  Using the angular diameter of the \ion{Si}{2}$^{*}$ emission derived from the observed line width, we can estimate the physical extent, $x_{SiII}$, of this region.  The physical diameter is then $x_{SiII}$~=~$d_{242}$$\tan$($\theta_{SiII}$), where $d_{242}$ is the distance to KISSR242.  Taking $H$~=~74.2~$\pm$~3.6 km s$^{-1}$ Mpc$^{-1}$~\citep{riess09} and $d_{242}$ (= $cz/H$)~=~153~$\pm$~8 Mpc, we find that 
$x_{SiII}$~=~252$^{+60}_{-55}$ pc.  

\subsection{Overlapping SSC \ion{H}{2} Regions}

The value we derive for the physical scale of the \ion{Si}{2}$^{*}$ emission clearly rules out an origin in \ion{H}{2} regions from 
individual massive stars (Str\"{o}mgen radius $x_{HII}$~$<$~1 pc) or even massive star-forming clusters (e.g. the Trapezium, $x_{HII}$~$<$~10 pc).  LCBGs are sites of intense star-formation, capable of producing nuclear starbursts of a similar scale as the emission we detect towards~KISSR242.  For example, the starburst core of M82 has a diameter of $\sim$~500 pc~\citep{smith05}, comparable to the inferred KISSR242 fine-structure line emitting region.  Perhaps on global scales, the overlap of \ion{H}{2} regions produced by individual SSCs (``Super \ion{H}{2}'' Regions?) can produce these nebular emission lines?  Existing observations argue that this is not the case.  
Far-UV fine structure emission has not been detected in any galactic star-forming region that we are aware of, nor in surveys of local starburst galaxies (with $IUE$; Kinney et al. 1993 or $HST$; Leitherer et al. 2010)\nocite{kinney93,leitherer10}.  Similarly, $FUSE$-based surveys of star-forming regions and starburst galaxies~\citep{keel04,grimes09} do not show low ionization fine-structure emission, even though strong emission lines from the (presumably) more abundant C$^{+}$ and N$^{+}$ ions reside in the 1030~--~1090~\AA\ $FUSE$ band.  

%Finally, this spectroscopic constraint argues against a scenario where numerous individual \ion{H}{2} 
%regions fall within the COS aperture, creating an ``effective'' source size that does not represent the
%actual distribution of the \ion{Si}{2}$^{*}$ emitting region.  

The detection of fine-structure emission in KISSR242 is not necessarily in conflict with existing observations of low-$z$ starburst galaxies.  COS samples a larger angular extent than most existing $HST$-GHRS/STIS observations, thus probing a greater spatial region of the target galaxies.  The sensitivity and resolution of COS improves the detection efficiency for fine-structure lines relative to previous wide-aperture UV spectrographs ($IUE$/HUT/$FUSE$).  
However, attempts at modeling the \ion{Si}{2}$^{*}$ emission in $z$~$\sim$~2~--~3 
LBGs as originating in photoionized \ion{H}{2} regions has proven largely unsuccessful.  Models with an incident UV radiation fields strong enough to excite the \ion{Si}{2} to the observed emission levels  
tend to overpredict other emission lines by as much as an order of magnitude~\citep{shapley03,erb10}.  

%\ion{Si}{2}$^{*}$ $\lambda$1533.43~\AA\ was observed in the \ion{H}{2} region surrounding %the bright B-star system $\alpha$ Vir~\citep{park10}, but again photoionization models do %not provide an unambiguous fit to the observations.

\subsection{Circumgalactic Gas Halo}

A second possibility is that the fine-structure emission we observe in KISSR242 originates in a static low-ionization circumgalactic halo.
This warm ($T$~$\sim$~3~$\times$~10$^{4}$ K) halo, or distribution of halo clouds, would be approximately at rest with respect to the systemic velocity of the galaxy.  It could be populated via outflows~(e.g., \S3.2), or gas accretion during a recent merger event.  This latter scenario is attractive as it explains the blue ``plume'' seen in the optical (Figure 1), presents a fuel source for the current epoch of intense star-formation in KISSR242, and even allows for the possibility that one of the resolved sources in the near-UV target acquisition image (Figure 1 inset) is a recently acquired additional nucleus.  If this picture of the \ion{Si}{2}$^{*}$ distribution is correct, the role of collisional excitation should be re-evaluated.  The critical density of the observed \ion{Si}{2}$^{*}$ lines is $n_{cr}$~$\approx$~2~$\times$~10$^{3}$ cm$^{-3}$ at these temperatures~\citep{bahcall68}.  This is characteristic of the intermediate density interstellar clouds that could be populating the halo via outflow or accretion.
This picture is in qualitative agreement with results showing that interactions and mergers are intimately tied to the star-formation in local LCBGs~\citep{barton01} and suggests that the \ion{Si}{2}$^{*}$ emission may be associated with the extended blue emission components of LCBGs~\citep{noeske06}. 
Further interpretation along this track would be speculative without additional observations, and a full simulation of the inflow/outflow dynamics of this system is beyond the scope of this Letter, which is meant to present new observational results for a low-$z$ LBG analog.  We refer the reader to~\citet{steidel10} for a kinematical discussion of LBGs at 2~$\lesssim$~$z$~$\lesssim$~3.

\section{Spectral Imagery of UV Emission from Circumgalactic Baryons}

The data do not support a firm conclusion about the origin of the warm, low-ionization emission from KISSR242, although
we favor the interpretation of emission from diffuse halo gas.  If our interpretation is correct, this would be the 
first detection of emission from warm, low-ionization gas surrounding a star-forming galaxy in the low-redshift universe, 
complementary to the recent discovery of \ion{C}{4} emitting filaments in the halo of M87/intracluster medium of Virgo~\citep{sparks09}.  
Assuming a uniform circular emitting geometry, the \ion{Si}{2}$^{*}$ flux levels we observe can be converted into a surface brightness, $B$, in line units (LU~$\equiv$~photons s$^{-1}$ cm$^{-2}$ sr$^{-1}$), $B_{i}$~=~(2.73~$\times$~10$^{18}$) $I_{i}$$\lambda_{i}$/$\theta_{SiII}^{2}$, where the index $i$ denotes the individual fine-structure line, $I$ is the line strength in erg cm$^{-2}$ s$^{-1}$, $\lambda$ is the wavelength in \AA, and $\theta_{SiII}$ is the angular diamter in arcseconds.  Taking the $^{2}S_{1/2}$~$\rightarrow$~$^{2}P_{3/2}$, $\lambda_{lab}$~=~1309.28~\AA\ transition as an example, we compute a line surface brightness of $B_{1309.28}$~$\sim$~3~$\times$~10$^{7}$ LU.  This corresponds to a count rate of 0.14 photons s$^{-1}$ within the COS aperture, despite the large number of LU.  
These results suggest that while future space missions designed to map baryons via their diffuse far-UV emissions~\citep{sembach09,kauffmann10} from cool (H$_{2}$; $T_{gas}$~$\sim$~500~--~5000~K), warm (\ion{Si}{2}$^{*}$; $T_{gas}$~$\sim$~10$^{4}$), and 
hot (\ion{C}{4} and \ion{O}{6}; $T_{gas}$~$\sim$~10$^{5-6}$ K) circumgalactic halos will not suffer from a dearth of targets, 
relatively large collecting areas (2~--~4m class) will be required to carry out surveys of the low-$z$ universe.

%Spectral maps of various phases across a range of galaxy environments will enable one to test models of 
%structure formation in cold dark matter halos in unprecedented detail.

\acknowledgments

We thank Brian Keeney for enjoyable and instructive discussions regarding feedback mechanisms in star-forming galaxies.  This work was support by NASA grants NNX08AC146 and NAS5-98043 to the University of Colorado at Boulder.

%%%%%%%%%%%%%%%%%%%%%%%%%%%BIBLIOGRAPHY%%%%%%%%%%%%%%%%%%

%\bibliography{ms}


%%%%%%%%%%%%%A PAGE OF FIGURE CAPTIONS, PER STEVE's FORMATTING ADVICE, 04/19/04%%%%%

%%%%%COPYING THE FIGURE INPUT FROM PAUL's IO TORUS PAPER, 7/16/03%%%%%

\begin{deluxetable}{lccccc}
\tabletypesize{\footnotesize}
\tablecaption{KISSR242 Fine-Structure Emission Lines}
\tablewidth{0pt}
\tablehead{
\colhead{Species} & \colhead{$\lambda_{rest}$} & \colhead{$\lambda_{lab}$}  & \colhead{Flux} 
& \colhead{FWHM}  &  \colhead{$v_{hel}$} \\ 
     & (\AA) & (\AA) & ($\times 10^{-16}$ ergs cm$^{-2}$ s$^{-1}$) & (km s$^{-1}$) & (km s$^{-1}$) }
%\tableline
\startdata
\ion{Si}{2}$^{*}$             & 1194.45 & 1194.50 & 9 $\pm$ 1     & 89 $\pm$ 9  & -13 $\pm$ 11 \\ 
\ion{Si}{2}$^{*}$             & 1197.42 & 1197.39  & 4 $\pm$ 1     & 63 $\pm$ 11  & 6 $\pm$ 11 \\ 
\ion{Si}{2}$^{*}$             & 1264.95 & 1265.00   & 6 $\pm$ 2    & 66 $\pm$ 13  & -12 $\pm$ 11 \\ 
\ion{Si}{2}$^{*}$             & 1309.29 & 1309.28   & 9 $\pm$ 1    & 82 $\pm$ 9  & 2 $\pm$ 11 \\ 
 \enddata
%% NOTES IN TABLE
\end{deluxetable}

%% The following command ends your manuscript. LaTeX will ignore any text
\end{document}